\documentclass{elsart}

\usepackage{natbib}

 \usepackage{graphics}
 \usepackage{graphicx}
 \usepackage{epsfig}

\usepackage{amssymb}

\begin{document}

\begin{frontmatter}



\title{Relativistic MHD Winds from Rotating Neutron Stars}


\author[l1]{Bucciantini, N.},\ead{nbucciantini@astro.berkeley.edu}
\author[l2]{Thompson, T.~A.},
\author[l1]{Arons, J.},
\author[l1]{Quataert, E.},
\author[l3]{Del Zanna, L.}

\address[l1]{Astronomy Department, University of Califonia at Berkeley, 601 Camblell Hall, Berkeley, CA, 94706, USA}
\address[l2]{Department of Astrophysical Sciences, Peyton Hall, Ivy Lane, Princeton, NJ, 08544, USA}
\address[l3]{Dipartimento di Astronomia, Universit\`a di Firenze, L.go E.Fermi 5, 50125, Firenze, Italy}

\begin{abstract}
% Text of abstract
We review here recent numerical results concerning the acceleration and collimation of relativistic outflows from neutron stars, and we discuss their implications for models of magnetars in their hypothesized initial high rotation states. New results with different injection conditions and extending to much larger distance from the central star are also presented and discussed, showing that, only in the case of weakly magnetized winds, collimated outflows are indeed possible. We finally comment on the possibility that newly born magnetars might trigger GRBs, and we show that numerical results do not support such hypothesis.
\end{abstract}


\end{frontmatter}

% main text
\section{Introduction}
\label{sec:intro}

The increasing evidence for SN-GRB association, suggests that the properties of the relativistic outflow associated with the burst might be related to the formation of a compact remnant, either a black hole or a neutron star at the center of the progenitor. The most promising model at the moment invokes the presence of a high density accretion disk around a central black hole \citep{macfad99}, however an alternative model involving the presence of a rapidly rotating magnetar has been presented \citep{usov92,thom04}.

If magnetars are born with millisecond period, their rotational energy ($\sim 10^{52}$ ergs) can be much larger than the energy of the supernova explosion ($\sim 10^{51}$ ergs), and because of their strong magnetic field, the spin-down time can be extremely short. They can in principle release most of their energy in a few minutes, a time shorter than the dynamical time of the surrounding supernova envelope, as required for a GRB. However to properly assess the validity of this model, one must also verify if the conditions of the outflow agree with the observed properties of GRBs, in particular the acceleration and collimation properties of the wind.

The initial spin period for neutron stars, is not well constrained. \citet{lai01}, analyzing the spin-kick alignment in pulsars, were able to put an upper limit to the initial period in the range 0.1-0.01 seconds, based on the kick mechanism. A lower limit on the period can be placed considering energy loss by gravitational waves. \citet{ho00} show that gravitational waves can be competitive with electromagnetic losses for periods shorter than 0.01 seconds, and a magnetic field weaker than $10^{14}$ G. However, newly born magnetars might have a higher magnetic field, and as we will show, mass loading, by opening more magnetic lines, can increase the wind losses well above the dipole prediction (closer to monopole).  

After the core-collapse, the newly born magnetar is hot (sound speed at the surface  $\sim 0.1-0.03 c$ ), and has an ``extended atmosphere''. This proto-magnetar cools and contracts on timescale, $\tau_{\rm KH}\sim10-100$\,s, radiating its gravitational binding energy ($\sim$$10^{53}$ ergs) in neutrinos of all species
\citep{pons99}.  The cooling epoch is accompanied by a thermal wind, driven by neutrino energy deposition, which emerges into the post-supernova-shock ejecta  \citep{thom01}. As the neutrino luminosity decreases, the thermal pressure at the edge of the proto-neutron star surface decreases, and the wind region becomes increasingly magnetically dominated, so that the outflow must eventually become relativistic. For magnetar-strength surface fields ($B_0\sim 10^{15}$ G) the magnetic field dominates the wind dynamics from just a few seconds after the core-collapse \citep{thom03}.

Observations clearly show that ultrarelativistic outflows are produced by pulsars, however theoretical models of the acceleration of such winds have proved until now to be unable to explain the observations. In the simple 1D radial model for relativistic winds from compact rotators by \citet{michel69}, solutions can be parametrized in terms of $\sigma=\Omega^2\Phi^2/\dot{M}$, where $\Omega$ is the rotation rate, $\Phi$ the magnetic flux and $\dot{M}$ the mass flux. $\sigma$ is the maximum Lorentz factor the wind can achieve if all magnetic energy is converted into kinetic energy, and at the base of a pulsar it is $\sim 10^6$. The results by \citet{michel69} show that the asymptotic Lorentz factor tends to $\sigma^{1/3}$. This is known as the $\sigma-\gamma$ problem \citep{arons04}. If the flow is not perfectly radial then acceleration can be more efficient \citep{beg94}, although acceleration to four-velocities in excess of $\sigma^{1/3}$ are quite slow, with $\gamma$ increasing only logarithmically with radius. The presence of a magnetic field can cause collimation of the flow along the rotation axis, and divergence along the equator, so in principle one might expect magnetic collimation, wind acceleration and jet formation to be closely related. However when the flow speed reaches values close to $c$ the Coulomb term in the electromagnetic force cannot be neglected and it balances the magnetic hoop stress. Solutions of the relativistic Grad-Shafranov equation for a monopolar field in the trans-fast regime \citep{beskin98} have shown that acceleration is too slow and the asymptotic results of the Michel model apply also in 2D axisymmetric geometry, meaning that the flow remains almost radial.

Past investigation has focused mostly on the pulsar regime ($\sigma \sim 10^3 - 10^6$), when the wind is almost force-free. However, in the case of newly born magnetar, the outflow is expected to evolve from a low-$\sigma$ regime to a high-$\sigma$ regime, and the details of such evolution might be important in the modeling of its interaction with the progenitor's envelope. Moreover it is not clear for what value of $\sigma$ the wind decollimates, and when the $\gamma\propto\sigma^{1/3}$ regime is achieved. Finally contrary to the case of pulsars, where injection is super-slow and the mass loss rate depends on the details of gap physics, for newly born magnetars, the wind is thermally driven at the base and the mass loss rate depends on the neutrino luminosity. As a consequence one might try to estimate how much time after the collapse the wind becomes relativistic, and compare the derived  energetic and timescales with observed values.

\section{Outflow properties of the wind}
\label{sec:res}

A numerical study of the outflow properties in the case of a pressure driven wind for a regime representative of the first few seconds after the formation of a proto-magnetar, has been presented by \citet{me06}, to which the reader is referred for a detailed description of the algorithm and of the settings. Here we will also discuss new results that have been obtained for different injection conditions and that extend to a much larger distance from the neutron star.

Simulations are performed on a 2D spherical grid in Schwarzschild metric, with 100 uniform zones in the $\theta$ direction, and a radial logarithmic grid with 200 zones per decade. The neutron star is assumed to be 10 km in radius and 1.4 solar mass. At the surface of the star we fix the radial component of the magnetic field, density and sound speed $c_s=0.1c$. In the case of supersonic injection we fix also the radial component of the velocity. For a newly born magnetar, injection is subsonic and is found self-consistently in the simulation. In order to reproduce neutrino heating within an ideal gas equation of state, we assumed an adiabatic coefficient $\Gamma=1.1$. Different values of $\sigma$ are obtained by changing the ratio $B^2/\rho c^2$ at the base of the star.

In analogy with the 1D model, solutions in the 2D axisymmetric monopole case can be parametrized by the average value of $\sigma$ over all streamlines. In the case of a dipolar magnetic field, despite the presence of a closed zone, and the fact that within the Light Cylinder the field structure is still close to a dipole, we found that solutions can again be parametrized in term of $\sigma$, now defined as an average done only on open streamlines.

Both in the case of a dipolar and monopolar magnetic field, solutions show that for $\sigma<1$, the energy and angular momentum losses at large distances are directed along the rotation axis. The outflow velocity tends to be higher at intermediate latitudes, contrary to the simple expectation that magnetocentrifugal acceleration should be more efficient along the equator. The flow is however almost thermally driven and the asymptotic speed non relativistic. One expect that as $\sigma$ goes to zero, there should be no collimation (the Alfvenic surface moves inside the sonic surface). On the other side, if $\sigma$ is too high, relativistic effects decollimate the flow. This suggests that there should be an optimum value of $\sigma$. Numerical results, for our specific problem, suggest that this optimum value is between $\sigma=0.01$ and $\sigma=0.5$, clearly in the non-relativistic regime.

As soon as $\sigma$ exceeds $10$ the flow becomes almost radial, the energy and angular momentum fluxes rapidly approach the force-free values. Even if $v << c$ inside the Light Cylinder, asymptotic results agree with the flow structure found by \citet{bog01}. Energy flux scales as $\sin^2{(\theta)}$, and  $\gamma\propto\sin{(\theta)}$ within 10\% accuracy, in agreement with the solution of the exact monopole. However within our computational domain the terminal $\gamma$ is $< \sigma$, pointing to inefficient acceleration. Mass flux is also shaped due to the change in centrifugal support at different latitudes. For the monopole case, on the equator it appears to be 5 times higher than in the non rotating case.

The analytic theory by \citet{beskin98}, in the case of uniform mass loading and high $\sigma$, gives the following relation between the Lorentz factor at the fast surface $\gamma_f$, and the radius of the fast surface $r_f$, close to the equator:
\begin{equation}
\gamma_f/r_f\propto \sin(\theta).
\end{equation}
We find from a numerical solution with $\sigma \sim 200$ that $\gamma_f/r_f\propto \sin(\theta)^{0.8}$, which agree with the theoretical expectation within 5\% error a $45^\circ$. It is not clear if this small difference is due to simplification in the perturbative approach by \citet{beskin98}, as already pointed out by \citet{bog01}, if it is due to a different mass loading, or to the fact that our value of $\sigma$ might not be high enough. 

In the case of a dipole, current sheets are present both at the edge of the closed zone and at the equator, numerical resistivity is important, limiting our investigation to cases with $\sigma < 20$. Even if the asymptotic structure is close to force-free, contrary to expectations from force-free models \citep{cont99}, we find that the location of the Y-point at these intermediate value of $\sigma$ is well inside the Light Cylinder. This is due to the presence of a finite pressure plasma at the base of the star. In this case the extent of the closed zone is limited by the pressure equilibrium at the Y-point. We find that the Alfvenic surface rapidly approaches the Light Cylinder, except very close to the equator, where it tends to be closer to the star. In Ideal MHD one expect the Alfvenic surface at the equator to recede to the location of the Y-point, but in our case, due to numerical resistivity on the equator itself this does not happen. Even if locally, at the equator, the Alfvenic surface is inside the Light Cylinder, this does not affect the global properties of the wind, which depend on the average location of the Alfvenic surface. 

We are able to relate the amount of open magnetic flux (which can be used to parametrize energy losses), to the location of the Y-point, $R_Y$:
\begin{equation}
\Phi_{open}=4\pi R_{NS}^2 \left(\frac{3}{5}B_{dip}\frac{R_{NS}}{R_Y}\right),
\end{equation}
where $R_{NS}$ is the radius of the neutron star, and $B_{dip}$ is the equatorial strength of the dipolar magnetic field at the stellar surface. Extrapolating to the force-free limit, our result is in agreement with \citet{gruz05}. We also verified that the ratio of the Y-point radius to the Light Cylinder $R_Y/R_{L}$ drops with decreasing $\Omega$, as one would expect for a braking index less than 3.

However, due to numerical constraints, those simulations were limited to a region extending just 100-200 times the radius of the neutron star, much smaller than the typical size of the region where the interaction with the progenitor envelope is supposed to take place, thus offering only an indication of the possible outcome. Also, the role played by the equation of state remained unclear. In order to model a thermally driven wind an almost isothermal equation of state was chosen, but as the neutrino luminosity drops, the wind becomes more and more adiabatic, and the effects of neutrino heating remain confined to the surface of the proto-neutron star. Finally, it was important to verify how robust was the result concerning the size of the closed zone, if gravity was important, and how different injection conditions could have affected the final state of the wind.

We therefore performed a new series of simulations, for the dipolar magnetic field, by using different injection conditions. We assumed a polytropic wind with $\Gamma=4/3$, low pressure $p/\rho c^2 \sim 0.01$, and we fixed the value of density, pressure, radial component of the magnetic field and poloidal velocity at the surface of the star. We assume the injection speed to be super-slow, $V_{inj}=0.4c$, along open field lines and zero on closed field lines. In this case the mass loss rate is imposed as a boundary condition, and one can neglect the effect of gravity. We also assume that the mass loss rate is independent of latitude. This is a reasonable approximation, in a dipolar case, where most of the mass comes from a small polar region (centrifugal support is not important). We found that the same constraint on the maximum value of sigma that could be achieved in the simulations, still hold even for this different settings.

Let us first compare these new results with the previous simulations. In discussing the evolution of a newly born magnetar, a key element is their spin down, and this is related to the size of the closed zone. Our results show that for a given value of sigma, in the absence of gravity, the size of the closed zone tends to be slightly larger. However the difference is minimal and can be neglected to a first order approximation. In \citet{me06} we suggested that the location of the Y-point, in the regime of interest, was governed by pressure equilibrium, at the Y-point itself, however we were not able to efficiently change the pressure at the base of the closed zone to verify such assertion. We can now confirm that this is indeed the case. We verify that by increasing for example the thermal pressure at the base of the closed zone the size of the close zone itself diminishes. This suggests that a model like \citet{mestel} could be used to relate the temperature at the base of the atmosphere (and so the neutrino luminosity) to the location of the Y-point and the spin-down power. However substantial variations in the location of the Y-point, require large pressure variations. The reason is that, at the temperatures typical of newly born magnetars, the radial pressure profile in the closed zone is quite shallow, while the confining magnetic pressure in the outer region $B^2\propto r^{-6}$. This is the reason why we recover similar results with or without gravity.

Another important result found in previous simulations was that, once normalized in term of $\sigma$ evaluated on open field lines, the energy and angular momentum losses of the dipole scaled in the same way as the monopole. While this is a reasonable expectation in a force-free regime, the result applied also to mass loaded cases. The new results show that energy and angular momentum losses, still have the same scaling with $\sigma$, with errors less than 15\%. It appears that, despite very different injection conditions, a different adiabatic coefficient for the gas, and the absence of gravity, as long as $\sigma > 1$, global properties are unchanged.

Finally the use of superslow injection conditions allows us to model the structure of the wind to larger distances. In previous simulations it was pointed out that the fast-magnetosonic surface, was not inside the computational domain along the axis. However we show that little variation was expected. By using superslow injection and a fixed mass flow rate, we are able to resolve the fast-magnetosonic surface inside the computational box. This guarantees that the boundaries do not exert any torque on the magnetosphere. One can thus extend the solution, to larger distances. Solutions were extended up to a distance of 10000 $R_{NS}$, comparable with the size of the collapsed core region. Given the fact that we are not able at the moment to investigate cases with $\sigma >30$, at larger distances a certain degree of collimation and acceleration is achieved. We verified that for $\sigma<5$, the flow does collimate along the axis, and the energy flux is higher at higher latitudes. Moreover very close to the axis, a very narrow channel is formed, which is unresolved in our grid. Such channel however does not contain a significant fraction of the energy flux. Acceleration is in this case efficient, and the flow reaches equipartition, even if the terminal Lorentz factor is small. For values of $\sigma\sim 5$, the energy distribution in the wind appears to be more isotropic. For values of $\sigma > 15$, even at such large distances, the energy flux is mostly directed along the equatorial plane. Acceleration is less efficient. Theory states that the acceleration should be logarithmic, but the low value of sigma we used does not allow us to verify this logarithmic trend. Even if the wind is still magnetically dominated, the terminal Lorentz factor $\gamma$ appears to be higher than the quasi force-free predicion $\sigma^{1/3}$. We find that in the case $\sigma \sim 25$ the Lorentz factor seems to saturate at $\gamma \sim 8$.

\citet{mel96} suggested that at large distance from the Light Cylinder MHD might be violated due to ``current starvation''. The distance at which this happen is still purely constrained, depending on the wind structure, current distribution and pair production, but is in general thought to be for standard pulsars greater than $10^5$ times the Light Cylinder radius. Soon after birth, mass loading is several orders of magnitude higher than what gap physics predict in regular pulsar, so that MHD will hold for very large distances, well outside the core of the progenitor.

\section{Conclusion}
\label{sec:concl}

Regarding the general problem of relativistic outflows, our recent numerical results have confirmed that the acceleration of relativistic winds from neutron stars is not efficient, as predicted by simplified theories. The wind remains magnetically dominated up to large distances from the star, the flow structure rapidly approaches the expectation from the exact monopole solution, as soon as $\sigma>10$, there is no evidence for the formation of a collimated energetic jet at high $\sigma$. The use of a more realistic dipolar field does not seem to change significantly the wind properties far from the Light Cylinder. On the contrary the energy and angular momentum losses seem to depend only on the integrated value of $\sigma$, and, for $\sigma>10$, on the amount of open magnetic flux. These results appear not to be sensitive to the specific injection conditions, or to the equation of state used for the plasma. 

Regarding the possibility that a magnetar could trigger a GRB, we conclude that, in ideal MHD, acceleration to relativistic speeds requires very high $\sigma$, much higher than what is thought to characterize newly-born magnetars. Moreover there is no evidence of energy collimation at high magnetization. However non-ideal effects might be important in this intermediate-$\sigma$ regime. If the star is not an aligned rotator, a striped wind region will form. Dissipation of the current sheet, in a striped wind, could be invoked to provide acceleration. However it is not clear what would be the efficiency of such process, mostly because of uncertainties in the value of resistivity. Even if non ideal effect might be invoked to provide the required acceleration, it is not obvious that a jet can form, it is more likely that the energy will still be concentrated in the equatorial plane, as is suggested by X-ray images of Pulsar Wind Nebulae. In this case one would naively expect a sheet like structure, instead of a beam, which is not consistent with observation of GRBs. 

The fact that the Y-point is interior to the light cylinder (and the Alfven radius), has also important consequences in the spin-down evolution of the neutron star. In order to have a GRB, a magnetar must retain most of its rotational energy until the neutrino luminosity has dropped enough to achieve $\sigma\sim 100-300$. However a smaller closed zone imply more open flux, and higher spin-down luminosity. As a consequence most of the rotational energy of the star could be lost in few seconds, and not enough power might be available at later times  when the outflow becomes relativistic.

Recent models for GRBs have focused on the details of the interaction between a relativistic outflows from a central compact source and the surrounding envelope of the progenitor \citep[e.g.][]{uzd06}. Results suggest that the envelope plays a key role in the collimation of the outflows. It is thus important to consider, in the case of a magnetar wind, the effect of the surrounding medium. No detailed study of the interaction of a magnetar wind with the environment has been presented yet, however we can point to some elements that might be important for collimation. First, as shown by our results, at the beginning, when the neutrino luminosity is high, the wind might be collimated, even if non relativistic. This collimated wind can preshape an elongated cavity in the progenitor. At later times, when the wind becomes relativistic, the shape of the cavity might create a preferential direction along which the wind might break through the envelope. Moreover if some form of nebula is created by the wind inside the progenitor, the anisotropic pressure, due to a toroidal magnetic field, might also contribute to the overall elongation of the cavity. A more accurate investigation of such interaction will be carried on in the future.  

\section*{Acknowledgements}
This work has been supported by NSF grant AST-0507813, and by NASA grant NAG5-12031. NB was supported partly by NASA through Hubble Fellowship grant HST-HF-01193.01-A, awarded by the Space Telescope Science Institute, which is operated by the Association of Universities for Research in Astronomy, Inc., for NASA, under contract NAS 5-26555. 

% \appendix

% \section{}
% \label{}

% Bibliographic references with the natbib package:
% Parenthetical: \citep{Bai92} produces (Bailyn 1992).
% Textual: \citet{Bai95} produces Bailyn et al. (1995).
% An affix and part of a reference:
%   \citep[e.g.][Ch. 2]{Bar76}
%   produces (e.g. Barnes et al. 1976, Ch. 2).

\end{document}